\documentclass{article}

\usepackage{amsthm}
\usepackage[english]{babel}
\usepackage{amsfonts}
\usepackage{amsmath}
\usepackage{amssymb}
\usepackage{graphicx}
\usepackage{graphics}
\usepackage{caption}
\usepackage{subcaption}
\usepackage{dcolumn}
\usepackage{bm}
\usepackage[margin=3.8cm]{geometry}

\begin{document}

\title{Inverse of the Vandermonde and Vandermonde confluent matrices}

\author {H\'{e}ctor Moya-Cessa, Francisco Soto-Eguibar\\
             \small Instituto Nacional de Astrof\'{\i}sica, \'Optica y Electr\'onica, INAOE\\
             \small Calle Luis Enrique Erro No. 1\\
             \small Santa Mar\'{\i}a Tonantzintla, Puebla. 72810 Mexico\\
              \\
              \\
              Published as: \\
              Applied Mathematics and Information Sciences \textbf{5}(3) (2011), 361-366.}
\date{}

\maketitle

\begin{abstract}
The inverse of the Vandermonde and confluent Vandermonde
matrices are presented. In the case of the Vandermonde matrix, we
present a decomposition in three factors, one of them a diagonal
matrix. The evaluation of such inverse matrices is a key point to
find functions of a matrix, namely exponential functions
(evolution operators) and  logarithmic functions (entropies) in
quantum mechanical topics.\\

\end{abstract}

\section{Introduction}
Although Vandermonde systems arise in many approximation and
interpolation problems \cite{Gautschi1,Turner}, they also appear
when we need to solve systems of differential equations
\cite{Tou}, such as when we interact a multi-level atom with a
classical or quantum field, a  trapped ion with a laser field (see
for instance \cite{Abdel-Aty}), etc. When an atom interacts with a
quantized field they get entangled \cite{Moya}. This produces
that, by analyzing the density matrix of the atom or the density
matrix of the quantized field, we can determine when they
disentangle. To do so we need either to calculate the entropy of
the sub-systems and this requires  evaluation of  functions of
(density) matrices.\\
In the particular case of the entropy, we need
to calculate logarithmic functions of the sub-system's density
matrices \cite{Moya}. Vandermonde matrices, and in particular,
their inverse, are helpful to determine such functions. A more
common function is the exponential function of a matrix, as a
Hamiltonian may be written usually in matrix form, and therefore
the solution of Scr\"odinger equations involve the use of
evolution operators, i.e. exponentials of Hamiltonians (see for
instance \cite{Amaro}, also for the case when superoperators are
considered, and \cite{Guasti} for time dependent Hamiltonians).
The key point for the evalution of such functions is to find the
inverse of a Vandermonde matrix or of the confluent Vandermonde
matrix (in case there are repeated eigenvalues). The purpose of
the present paper is precisely this.

\section{Vandermonde matrices}
A matrix $N\times N$ of the form
\begin{equation}V=\left(
\begin{array}{cccccccc}
1 & 1 & 1 & . & . & . & 1 & 1 \\
\lambda _{1} & \lambda _{2} & \lambda _{3} & . & . & . & \lambda
_{N-1} &
\lambda _{N} \\
\lambda _{1}^{2} & \lambda _{2}^{2} & \lambda _{3}^{2} & . & . & .
& \lambda
_{N-1}^{2} & \lambda _{N}^{2} \\
\lambda _{1}^{3} & \lambda _{2}^{3} & \lambda _{3}^{3} & . & . & .
& \lambda
_{N-1}^{3} & \lambda _{N}^{3} \\
. & . & . & . & . & . & . & . \\
. & . & . & . & . & . & . & . \\
. & . & . & . & . & . & . & . \\
\lambda _{1}^{N-1} & \lambda _{2}^{N-1} & \lambda _{3}^{N-1} & . &
. & . &
\lambda _{N-1}^{N-1} & \lambda _{N}^{N-1}%
\end{array}%
\right)
\end{equation}
or
\begin{equation}
V_{i,j}=\lambda _{j}^{i-1} \qquad  i=1,2,3,...,N; \qquad
j=1,2,3,...,N
\end{equation}
is said to be a Vandermonde matrix \cite{Fielder,Gautschi1}.\\
The determinant of the Vandermonde matrix can be expressed as
$$\det \left( V\right) =\prod\limits_{1\leq i\leq j\leq N}\left(
\lambda _{j}-\lambda _{i}\right). $$
Therefore, if the numbers $\lambda _{1},\lambda _{2},...,\lambda _{N}$ are distinct, $V$ is
a nonsingular matrix \cite{Fielder}.\\
When two or more $\lambda _{i}$ are equal, the corresponding
matrix is singular. In that case, one may use a generalization
called confluent Vandermonde matrix \cite{Gautschi1,Wikipedia},
which makes the matrix non-singular, while retaining most
properties. If $\lambda _{i}=\lambda _{i+1}=...=\lambda _{i+k}$
and $\lambda _{i}\neq \lambda _{i-1}$, then the $ \left(
i+k\right) $th column is given by
\begin{equation}
C_{i+k,j}=\Big\{%
\begin{array}{cc}
0 & j\leq k \\
\dfrac{\left( j-1\right) !}{\left( j-k-1\right) !}x_{i}^{j-k-1} & j>k%
\end{array}%
\end{equation}
The confluent Vandermonde matrix looks as
\[
C=
\]
{\small
\begin{equation}
\left(
\begin{array}{ccccccccccc}
1 & 1 & ... & 1 & 0 &  0 & ... & 1 & 1 \\
\lambda _{1} & \lambda _{2} & ... & \lambda _{i} & 1 &  0 & ... &
\lambda _{m-1} & \lambda _{m} \\
\lambda _{1}^{2} & \lambda _{2}^{2} & ... & \lambda _{i}^{2} &
2\lambda _{i}
&  0 & ... & \lambda _{m-1}^{2} & \lambda _{m}^{2} \\
\lambda _{1}^{3} & \lambda _{2}^{3} & ... & \lambda _{i}^{3} &
3\lambda _{i}^{2} &  . & ... & \lambda _{m-1}^{3} & \lambda
_{m}^{3} \\
... & ... & ... & ... & ... &  \dfrac{\left( i-1\right) !}{\left(
i-k-1\right) !}\lambda _{i}^{i-k-1} & ... & ... & ... \\
... & ... & ... & ... & ... &  ... & ... & ... & ... \\
... & ... & ... & ... & ... &  ... & ... & ... & ... \\
\lambda _{1}^{n-1} & \lambda _{2}^{n-1} & ... & \lambda _{i}^{n-1}
& \left( n-1\right) \lambda _{i}^{n-2} &  \dfrac{\left( n-1\right)
!}{\left( n-k-1\right) !}\lambda
_{i}^{n-k-1} & ... & \lambda _{m-1}^{n-1} & \lambda _{m}^{n-1}%
\end{array}
\right) . \end{equation}}
Another way to write the $\left( i+k\right) $ column is using the
derivative, as follows
\begin{equation}C_{i,j+k}=\dfrac{dC_{i,j+k-1}}{d\lambda _{j}}.\end{equation}

\section{The inverse of the Vandermonde matrix}
In applications, a key role is played by the inverse of the
Vandermonde and confluent Vandermonde matrices \cite{Gautschi1,Turner,Tou,Oruc1,Oruc2,Gautschi 2, Wu,Yang}. Both matrices,
Vandermonde and confluent Vandermonde, can be factored into a
lower triangular matrix $L^{\prime }$
and an upper triangular matrix $U^{\prime }$ where $V$ or $C$ is equal to $%
L^{\prime }U^{\prime }$. The factorization is unique if no row or
column interchanges are made and if it is specified that the diagonal elements of $%
U^{\prime }$ are unity. \\
Then, we can write $V^{-1}=\left( U^{\prime }\right) ^{-1}\left(
L^{\prime }\right) ^{-1}$. Denoting $\left( U^{\prime }\right)
^{-1}$ as $U$, we have found that $U$ is an upper triangular
matrix whose elements are\\

$U_{i,j}=0$ if $i>j$

$U_{i,j}=\prod\limits_{k=1,\text{ }k\neq i}^{j}\dfrac{1}{\lambda
_{i}-\lambda _{k}}$ {\textrm{    otherwise}}.\\
\\
\\
The matrix $U$ can be decomposed as the product of a diagonal
matrix $D$ and other upper triangular matrix $W.$ It is very easy
to find that
\begin{equation}D_{i,j}=\Big\{
\begin{array}{cc}
\prod\limits_{k=1,\text{ }k\neq i}^{N}\dfrac{1}{\lambda
_{i}-\lambda _{k}} &
i=j \\
0 & i\neq j%
\end{array}%
\end{equation}
and
\begin{equation}W_{i,j}=\Big\{
\begin{array}{cc}
0 & i>j \\
\prod\limits_{k=j+1,\text{ }k\neq i}^{N}\left( \lambda
_{i}-\lambda
_{k}\right) & \text{otherwise}%
\end{array}.%
\end{equation}

\bigskip

The matrix $L=\left( L^{\prime }\right) ^{-1}$ is a lower
triangular matrix, whose elements are
\begin{equation}
L_{i,j}=\Big\{%
\begin{array}{cc}
0 & i<j \\
1 & i=j \\
L_{i-1,j-1}-L_{i-1,j}\lambda _{i-1} & i=2,3,...,N;\ j=2,3,...,i-1%
\end{array}.
\end{equation}
Summarizing, the inverse of the Vandermonde matrix can be written
as $V^{-1}=DWL$.

\section{The inverse of the confluent Vandermonde matrix}
We will treat now the case of the confluent Vandermonde matrix. We
suppose that just one of the values $\lambda _{i}$ is repeated, and it is repeated $%
m $ times. We make the usual LU decomposition, getting
$C=L_{c}^{\prime
}U_{c}^{\prime },$ where $L_{c}^{\prime }$ is a lower triangular matrix and $%
U_{c}^{\prime }$ an upper triangular matrix $U^{\prime }$. Then,
we can write $C^{-1}=\left( U_{c}^{\prime }\right) ^{-1}\left(
L_{c}^{\prime }\right) ^{-1}$. Denoting $\left( U_{c}^{\prime
}\right) ^{-1}$ as $U_{c}$, we have found that $U_{c}$ is an upper
triangular matrix whose elements are
\begin{equation}
\left( U_{c}\right) _{i,j}=0 \qquad i\qquad i>j
\end{equation}

\begin{equation}
\left( U_{c}\right) _{i,j}=\dfrac{\delta _{ij}}{\left( i-1\right) !} \qquad
i=1,2,3,...,m ;\qquad  j=1,2,3,...,m
\end{equation}

\begin{eqnarray}
\left( U_{c}\right) _{i,j}=-\dfrac{1}{\left( i-1\right)
!}\sum_{\alpha =m+1}^{j}\prod\limits_{\beta =i,\beta \neq \alpha
}^{j}\dfrac{1}{\left( \lambda _{\alpha }-\lambda _{\beta }\right)
}
\\
\nonumber
 i=1,2,3,...,m;
\\
\nonumber
 j=m+1,m+2,...,N
\end{eqnarray}

\begin{equation}
\left( U_{c}\right) _{i,j}=\prod\limits_{\beta =1,\beta \neq \alpha }^{j}%
\dfrac{1}{\left( \lambda _{i}-\lambda _{\beta }\right) }\qquad
i=m+1,m+2,...,N; \qquad j=i,...,N
\end{equation}
where it is understood that $\lambda _{m}=\lambda _{m-1}=...=\lambda
_{2}=\lambda _{1}$, and where the numbers $\lambda _{m},\ \lambda
_{m-1},\ ...,\ \lambda _{2}$ appear, they must be substituted by
$\lambda _{1}.$\\
The matrix $L_{c}=\left( L_{c}^{\prime }\right) ^{-1}$ is a lower
triangular matrix, whose elements are given by the following
recurrence relation,
\begin{equation}
( L_{c}) _{i,j}= \Big\{
\begin{array}{cc} 0 & if i<j
\\
1 & if i=j
\\
( L_{c}) _{i,ji-1,j-1}-( L_{c}) _{i,ji-1,j}\lambda _{i-1}& if
i=2,3,...,N;  j=2,3,...,i-1
\end{array},
\end{equation}
also here it is understood that $\lambda _{m}=\lambda
_{m-1}=...=\lambda _{2}=\lambda _{1}$, and where the numbers
$\lambda _{m},\ \lambda _{m-1},\ ...,\ \lambda _{2}$ appear, they
must be substituted by $\lambda _{1}.$\\
When more than one value is repeated, the inverse has blocks with
the same structure that we have already found.

\section{Conclusions} We have shown a form to determine the inverse
of Vandermonde and confluent Vandermonde matrices. Although
several studies exist for Vandermonde matrices, it is not so for
systems with repeated eigenvalues, which lead to confluent
matrices. Such inverse matrices are of importance in several
quantum mechanical topics where it is needed to find functions of
matrices, such as in quantum information processes, where
entropies play a key role.


\begin{thebibliography}{99}
\bibitem{Gautschi1} \textit{On inverses of Vandermonde and confluent Vandermonde matrices}. Walter Guatschi. Numerische Mathematik 4,(1962), 117-123.
\bibitem{Turner} \textit{Inverse of the Vandermonde matrix with applications} L. Richard Turner. NASA TN D-3547
\bibitem{Tou} \textit{Determination of the Inverse Vandermonde Matrix}. Julius T. Tou. IEEE Transactions on Automatic Control, Vol AC-9, Issue 3,  314 (1964).
\bibitem{Abdel-Aty} M. Abdel-Aty and H. Moya-Cessa, Phys. Lett. A,369, 372-376 (2007), {\it Sudden death and long-lived entanglement
of two trapped ions.}
\bibitem{Moya}   H. Moya-Cessa, L. Knight and A. Rosenhouse-Dantsker,  Phys. Rev. A 50, 1814-1821 (1994).
{\it Photon Amplification in a Two-Photon Lossless
Micromaser}
\bibitem{Amaro} L.M. Ar\'evalo-Aguilar, R. Ju\'arez-Amaro, J.M. Vargas-Mart\'{\i}nez, O. Aguilar-Loreto, and H. Moya-Cessa,
App. Math.  Inf. Sc. 2, 1, 43-49 (2008), {\it Solution of Master
Equations for the anharmonic oscillator interacting with a heat
bath and for parametric down conversion process.}
\bibitem{Guasti} M. Fern\'andez Guasti and H. Moya-Cessa, J. of Phys.  A {\bf 36}, 2069
(2003) {\it Solution of the Schroedinger equation for time
dependent 1D harmonic oscillators using the orthogonal function
invariant}; H. Moya-Cessa and M. Fern\'andez Guasti, Phys. Lett. A
{\bf 311}, 1 (2003) {\it Coherent states for the time dependent
harmonic oscillator: the step function}.
\bibitem{Fielder}
\textit{Special matrices and their applications in numerical
mathematics}. Miroslav Fiedler. 1986 by Martinus Nijhoff
Publishers and SNTL - Publishers of Technical Literature. ISBN
90-247-2957-2.
\bibitem{Wikipedia} http://en.wikipedia.org/wiki/Vandermonde\_matrix.
\bibitem{Oruc1} \textit{Explicit factorization of the Vandermonde matrix}.
Halil Oru\c{c}, George M. Phillips. Linear Algebra and its
Applications 315 (2000) 113--123.
\bibitem{Oruc2} \textit{LU factorization of the Vandermonde matrix and its
applications}. Halil Oru\c{c}. Applied Mathematics Letters 20
(2007) 982--987.
\bibitem{Gautschi2} \textit{On Inverses of Vandermonde and Confluent
Vandermonde Matrices III}. Walter Gautschi. Numer. Math. 29,
445-450 (1978).
\bibitem{Wu} \textit{On the Inverse of Vandermonde Matrix. }Sherman H. Wu.
IEEE Trans. Automat. Contr., vol. AC-11, p. 769, (1966).
\bibitem{Yang} \textit{On the LU factorization of the Vandermonde matrix. }%
Sheng-liangYang. Discrete Applied Mathematics 146 (2005) 102 --
105.
\end{thebibliography}
\end{document}